\documentclass[prc,aps,twocolumn,floats,floatfix,showpacs,nofootinbib,superscriptaddress]{revtex4}

\usepackage{amsmath}
\usepackage{amssymb}
\usepackage{amsfonts}
\usepackage{graphicx}
\usepackage{tabularx}
\usepackage{multirow}
\usepackage{lscape}
\usepackage{rotating}
\usepackage{pstricks}
\usepackage{version}

\def\be{\begin{equation}}
\def\ee{\end{equation}}
\def\bea{\begin{eqnarray}}
\def\eea{\end{eqnarray}}
\def\bean{\begin{eqnarray*}}
\def\eean{\end{eqnarray*}}
\newcommand{\bwt}{\begin{widetext}}
\newcommand{\ewt}{\end{widetext}}

\def\nn{\nonumber}
\def\nnn{\nonumber \\}

\newcommand{\eps}{\varepsilon}
\newcommand{\rhosat}{\rho_\mathrm{sat}}

\begin{document}


\title{Two-body contributions to the effective mass in nuclear effective interactions}

\author{D. Davesne}
\email{davesne@ipnl.in2p3.fr}
\affiliation{Univ Lyon, Universit{\'e} Claude Bernard Lyon 1, CNRS, IPNL, UMR 5822, 4 rue E.Fermi, F-69622 Villeurbanne Cedex, France}

\author{J. Navarro}
\email{navarro@ific.uv.es}
\affiliation{IFIC (CSIC-Universidad de Valencia), Apartado Postal 22085, 
             E-46.071-Valencia, Spain}

\author{J. Meyer}
\email{jmeyer@ipnl.in2p3.fr}
\affiliation{Univ Lyon, Universit{\'e} Claude Bernard Lyon 1, CNRS, IPNL, UMR 5822, 4 rue E.Fermi, F-69622 Villeurbanne Cedex, France}

\author{K. Bennaceur}
\email{bennaceur@ipnl.in2p3.fr}
\affiliation{Univ Lyon, Universit{\'e} Claude Bernard Lyon 1, CNRS, IPNL, UMR 5822, 4 rue E.Fermi, F-69622 Villeurbanne Cedex, France}

\author{A. Pastore}
\email{alessandro.pastore@york.ac.uk}
\affiliation{Department of Physics, University of York, Heslington, 
             York, Y010 5DD, United Kingdom}

\begin{abstract}
Starting from general expressions of well-chosen symmetric nuclear matter quantities derived for both zero- and finite-range effective theories, we derive some universal relations between them. We first show that, independently of the range, the two-body contribution is enough to describe correctly the saturation mechanism but gives an effective mass value around $m^*/m \simeq 0.4$ when the other properties of the saturation point are set near their generaly accepted values. Then, we show that a more elaborated interaction (by instance, an effective two-body density-dependent term on top of the pure two-body term) is needed to reach the accepted value $m^*/m \simeq 0.7-0.8$.
\end{abstract}


 
\date{\today}


\maketitle


\section{Introduction}

\label{sect:intro}

The penalty function used in the process of fitting a nuclear effective interaction (either zero- or
finite-range) is often built as a mixture of  experimental data on nuclei and some empirical values
concerning the equation of state of pure neutron matter and symmetric nuclear
matter (SNM) in the vicinity of the saturation point.
In particular, for SNM,  one  usually includes the Fermi momentum $k_F$,
the energy per particle $E/A$ at saturation, the compression modulus
$K_{\infty}$ and the effective mass at the Fermi surface $m^*/m$.

For both zero- and finite-range interactions, it is possible to derive simple expressions
for these SNM properties as functions of the interaction parameters. 
By  combining these functions in an appropriate way, we can eliminate some of the parameters and
get general relations which are convenient for the fitting process.
These simple analytical relations also show that some infinite nuclear matter properties 
might not be independent and, therefore, some constraints might be conflicting. 
For example, in the case of standard Skyrme interactions ({\it i.e.} zero-range interactions 
with momentum-dependent terms up to second order and one density-dependent contact term), the manifest correlation between the effective mass, 
the incompressibility and the power $\alpha$ of the density in the density-dependent term
has been already discussed in Ref.~\cite{cha97}.

In the present article, we investigate the origin of such a correlation, and other similar ones,  using different families of non-relativistic effective interactions. An essential point is to discern whether such a
correlation, or other similar ones, is general or is an artefact due to the
specific form of the adopted interaction.

By inspecting the scientific literature, we observe that when an effective interaction contains only two-body terms ({\it i.e.} no explicit three-body, four-body  or density-dependent terms) the effective mass in SNM is inevitably close to $m^*/m = 0.4$.
This value is obtained, for example, with the
SV~\cite{bei75} and SHZ2~\cite{sat12} Skyrme interactions, but also with the finite-range interaction B1~\cite{bri67} 
and the more recent class of regularized pseudo-potentials~\cite{ben17}.

The mechanism leading to a low effective mass for pure two-body interactions  was already identified years ago by Weisskopf~\cite{wei57}
(and more recently discussed by Nakatsukasa {\it et al.}~\cite{nak16}) and proved to be 
unavoidable for any interaction which gives a mean-field at most quadratic in momentum in SNM.
Our aim is thus to use these evidences as a starting point and to explore the possibility of finding
some general relations which can explain this obtained value for a generic two-body interaction, including finite-range ones.

The article is organised as follows: in Sect.~\ref{sec:2B} we focus on pure two-body terms and obtain some general relations in which we isolate the effective mass and give some numerical values from standard interactions. In Sect.~\ref{sec:DD} we extend the analysis by including an explicit two-body density dependent term. Finally, we expose our  conclusions in Sect.~\ref{sec:conclusion}.

%
%
\section{Two-body interaction}
\label{sec:2B}
%
%
We begin this section by defining some important  quantities which will be useful in the following. 
The starting point for our reasoning is the energy per particle $E/A$ in infinite symmetric nuclear matter.  This is a crucial quantity, since all the other relevant thermodynamical properties are related to $E/A$ via simple derivative operations.
We define the pressure as
\be
\label{therm-p}
P = \rho^2 \,\frac{\partial E/A}{\partial \rho}\;, 
\ee
and the isothermal compressibility
\be
\label{iso-K}
\frac{1}{\kappa_T} = \frac{\partial P}{\partial \rho}\;,
\ee
where $\rho$ is the nucleon (scalar-isoscalar) density in SNM.
For historical reasons, instead of Eq.~(\ref{iso-K}), it is traditional in nuclear physics to consider
 the compression modulus at saturation defined as
\be
\label{compress-0}
K_{\infty} = 9 \left( \rho^2\, \frac{\partial^2 E/A}{\partial \rho^2}
\right)_{\rho=\rho_0}\,.
\ee
Using Eq.~(\ref{therm-p}) this can be equivalently written as
\be
\label{compress}
K_{\infty} =  9 \left( \frac{1}{\kappa_T} - 2 \,\frac{P}{\rho}
\right)_{\rho=\rho_0}\,.
\ee
The consistency with Eq.~(\ref{compress-0}) is ensured by the fact that,
at saturation density $\rho_0$ one has $P=0$. 

Notice that any contribution to $E/A$ linear in the density $\rho_0$ gives a
quadratic contribution to $P$, with the same coefficient. 
Therefore, these two contributions cancel out exactly in the difference entering 
the right hand side of Eq.~(\ref{compress}).
This applies for instance to the $t_0$ term of the Skyrme interaction, and this is why the usual 
expression for $K_{\infty}$ found in the literature does not contain an explicit dependence on $t_0$. 
In the case of a finite-range momentum-independent interaction, since the contribution of the direct 
term to $E/A$ is also linear in $\rho_0$, the same conclusion applies. 
Based on this observation, we will adopt for the energy per particle at saturation the following 
expression
\be
\label{e0}
{\cal E}_0 = \left( \frac{E}{A} - \frac{1}{\rho} P \right)_{\rho=\rho_0}\;.
\ee
As discussed previously, such an expression does not dependent on $t_0$ in the case of a Skyrme interaction or on the 
direct term in the case of a finite-range momentum-independent two-body interaction.

%
%
\subsection{Weisskopf's relation}
\label{sec:weiss}
%
%

Years ago, Weisskopf~\cite{wei57} showed that the mean-field in nuclear matter should be momentum dependent. 
Assuming a quadratic dependence, he got a relation between the effective mass, the binding energy per particle 
and the Fermi momentum. 
For the sake of clarity, we derive here this relation with a modified notation.

Within the hypothesis of a quadratic momentum dependence, the mean field $U_i$ for a state $i$ with momentum $p_i$ 
can be written as
\be
U_i = U_0 + \frac{p_i^2}{p_F^2} \,U_1\;,
\label{U-weiss}
\ee
where $U_0$ and $U_1$ are some constants and $p_F=\hbar k_F$ the Fermi momentum.
The effective mass, defined through the relation
\be
\frac{p_i^2}{2m} + U_i \equiv \frac{p_i^2}{2m^*} + U_0\,,
\ee
can therefore be written as
\be
\frac{m}{m^*} = 1 + \frac{U_1}{{\varepsilon}_F}\;,
\label{weiss-1}
\ee
where ${\varepsilon}_F= \hbar^2 k_F^2 / 2m$ is the Fermi energy. 

As an illustration, we can mention that the above hypothesis embraces the standard Skyrme interaction: 
dropping here the density-dependent term (see Sec.~\ref{sec:DD} for a detailed discussion of this term), 
it is straightforward to show that the mean field can be written as
\be
U(k) = \frac{3}{4}\, t_0 \,\rho + \frac{3}{80}\, C_1^{(2)} \rho k_F^2
 +  \frac{1}{16}\, C_1^{(2)} \rho k^2 \;,
\ee
where $C_1^{(2)} = 3 t_1 + (5+4x_2) t_2$. 
Comparing with Eq.~(\ref{U-weiss}) we identify 
\be
U_1 = \frac{1}{16}\, C_1^{(2)} \rho k_F^2\;,
\ee
so that Eq.~(\ref{weiss-1}) leads to the familiar expression for the effective mass
\be
\frac{m^*}{m} =
  \left[1+ \frac{1}{8}\, \frac{m}{\hbar^2} \,C_1^{(2)} \rho \right]^{-1}\;,
\ee
for Skyrme interactions.

Weisskopf  established an interesting relation between $m^*/m$, 
$E/A$ and $\varepsilon_F$.
To rederive it, we start from the energy per particle written as (the brackets indicate an average 
over the particles states)
\bea
E/A & \equiv & \langle T\rangle + \frac{1}{2} \langle U \rangle \nnn
   & = & \frac{3}{5}\,{\varepsilon}_F + \frac{1}{2}\,U_0 + \frac{3}{10}\,U_1\,.
\eea
The separation energy of a particle at the Fermi surface is given by
\be
S_F = -( {\varepsilon}_F + U_F ) = - \left( {\varepsilon}_F + U_0 + U_1 \right)\;,
\ee
and it is related to the energy per particle via Hugenholtz-Van Hove theorem~\cite{hvh} with $S_F = -{\cal E}_0$ , so that one immediately 
gets~\footnote{We indicate in passing that Ref.~\cite{wei57} contains two misprints in the sign of $U_1$.}
\be
U_1 =  \frac{1}{2} \left( -5 {\cal E}_0  + {\varepsilon}_F \right).
\ee
Finally, the effective mass can therefore be written as
\be
\frac{m}{m^*} = \frac{3}{2} - \frac{5}{2}\,\frac{{\cal E}_0}{{\varepsilon}_F}.
\label{weiss-2}
\ee
For the commonly accepted values of ${\cal E}_0$ and $k_F$ ({\em i.e.}
$-16$~MeV and
$1.33$~fm$^{-1}$ respectively), the above equation leads to an effective mass 
$m^*/m \simeq 0.4$. 
Due to the assumed quadratic momentum dependence of the mean field, this result can only be used 
for a standard Skyrme interaction with no density-dependent term. 
In order to compare with a practical case, we thus consider the SV interaction~\cite{bei75} 
which is one of the very few Skyrme interactions with no density dependence.
The relevant SNM properties of SV 
interaction leads to the values ${\cal E}_0=-16.05$~MeV, $k_F=1.32$~fm$^{-1}$ and
$m^*/m=0.38$. 
Consistently, the use of Eq.~(\ref{weiss-2}) gives $m^*/m=0.38$.

Finally, let us mention that, as discussed by Weisskopf, the mean-field potential may have a 
momentum-dependence beyond the quadratic one. 
Such a dependence is examined below.

%
%
\subsection{Zero-range N3LO Skyrme interaction}
\label{sec:N3LO}
%
%

The central part of the Skyrme N3LO pseudo-potential~\cite{rai11,dany14}  with no additional density-dependent term reads
\begin{eqnarray} 
\label{eq:N3LO}
V_{\rm N3LO}^{c} &=& t_0 (1+x_0 P_{\sigma}) + \frac{1}{2} t^{(2)}_1 (1+x^{(2)}_1 P_{\sigma}) ({\bf k}^2 + {\bf k'}^2) \nnn
&+& t^{(2)}_2 (1+x^{(2)}_2 P_{\sigma}) ({\bf k} \cdot {\bf k'})  \nnn
&+& \frac{1}{4} t_1^{(4)} (1+x_1^{(4)} P_{\sigma}) \left[({\bf k}^2 + {\bf k'}^2)^2 + 4 ({\bf k'} \cdot {\bf k})^2\right] \nnn
&+& t_2^{(4)} (1+x_2^{(4)} P_{\sigma}) ({\bf k'} \cdot {\bf k}) ({\bf k}^2 + {\bf k'}^2) \nnn
&+& \frac{1}{2} t_1^{(6)} (1+x_1^{(6)}P_\sigma) ({\bf k}^2 + {\bf k'}^2) \nnn
&& \quad \times \left[({\bf k}^2 + {\bf k'}^2)^2 + 12 ({\bf k'} \cdot {\bf k})^2\right] \nnn
&+& t_2^{(6)}(1+x_2^{(6)}P_\sigma)
({\bf k'} \cdot {\bf k}) \nnn
&& \quad \times \left[3({\bf k} ^2 + {\bf k'}^2)^2 +4({\bf k'} \cdot {\bf k})^2\right].
\end{eqnarray}
where a $\delta({\bf r}_1-{\bf r}_2)$ function factorizing all terms is to be
understood, but has been omitted for the sake of clarity.  
From this pseudo-potential,  we compute the energy per particle as
\bea
\frac{E}{A} & = & \frac{3}{10}\,\frac{\hbar^2}{m}\,c_s\,\rho^{2/3}
                + \frac{3}{8}   \,t_0  \rho
                + \frac{3}{80}  \, C_1^{(2)} \, c_s   \, \rho^{5/3}   \nnn
            & + & \frac{9}{280} \, C_1^{(4)} \, c_s^2 \, \rho^{7/3} 
                + \frac{2}{15}  \, C_1^{(6)} \, c_s^3 \, \rho^{9/3} \;,
\label{E-N3LO} 
\eea
where we have defined $c_s= (3 \pi^2/2)^{2/3}$. 
The three quantities we are interested in are
\begin{align}
{\cal E}_0 = \frac{1}{5}\, \varepsilon_F &- \frac{1}{40}\, C_1^{(2)}  \rho_0 k_F^2  \nonumber \\
& - \frac{3}{70} \, C_1^{(4)} \, \rho_0 \, k_F^4
  - \frac{4}{15} \, C_1^{(6)} \, \rho_0 \, k_F^6  \, ,  
\label{e0-N3LO}
\end{align}
\begin{align}
K_{\infty}  = - \frac{6}{5}\, \varepsilon_F
& +  \frac{3}{8}\, C_1^{(2)} \rho_0 k_F^2 \nonumber \\
& + \frac{9}{10}\, C_1^{(4)}   \rho_0 k_F^4
  + \frac{36}{5}\, C_1^{(6)}  \rho_0 k_F^6  \, , 
  \label{k-N3LO} 
\end{align}
and
\begin{align}
\frac{\hbar^2}{2m^*} k_F^2 = \varepsilon_F
& +\frac{1}{16}\, C_1^{(2)} \rho_0 k_F^2\nonumber \\
& + \frac{1}{8}\, C_1^{(4)} \rho_0 k_F^4
  + \frac{9}{10}\, C_1^{(6)} \rho_0 k_F^6 \, . 
  \label{eff-N3LO}
\end{align}
The coupling constants  entering these expressions are related to the parameters of the pseudo-potential in Eq.~(\ref{eq:N3LO}) as 
$C_1^{(n)} = 3 t_1^{(n)} + (5+4x_2^{(n)}) \, t_2^{(n)}$.  

In the case of a standard Skyrme interaction (corresponding to N1LO), the
three above expressions depend only on the coupling constant labeled $C_1^{(2)}$. 
This means that these quantities are correlated by pairs.
Besides  Eq.~(\ref{weiss-2}) we can obtain two more relations
\begin{eqnarray}
15 {\cal E}_0 + K_{\infty} &=& \frac{9}{5} \varepsilon_F\;, \\
m/m^*&=&\frac{6}{5}+\frac{K_{\infty}}{6 \varepsilon_F}\;.
\end{eqnarray}
Such relations are clearly specifically related to the form of the interaction 
({\it i.e.} only valid for SV-like interactions) and do not hold in any other cases. 

Using Eqs.~(\ref{e0-N3LO}-\ref{eff-N3LO}) we derive an expression for the effective mass where we exactly cancel out both coefficients $C_1^{(2)}$ and $C_1^{(4)}$  by taking the following combination
\be
 \frac{35}{24}\, {\cal E}_0 - \frac{5}{72}\, K_{\infty}
       + \frac{\hbar^2}{2m^*}\, k_F^2 \,.
 \label{magic}
\ee
It leads to the following relation
\bea
\frac{m}{m^*} \, = \, \frac{11}{8} \, + \, \frac{5}{72} 
                   \, \frac{K_{\infty} - 21 {\cal E}_0}{\varepsilon_F}  
              \, + \, \frac{1}{90} \, \frac{1}{\varepsilon_F} \, C_1^{(6)} \, \rho_0 \, k_F^6  \, .
\label{rel-N3LO}
\eea
This is an interesting result: keeping in mind that the coefficients $C_1^{(n)}$
are related to terms simulating finite-range effects in the N3LO Skyrme
interaction~\cite{dav16}
this equation states that finite-range effects only appear at
third order (N3LO) and thus constitutes an exact result for N1LO and N2LO interactions.

In order to have a quantitative insight of the effect beyond N2LO, we have
displayed in Table~\ref{tab:2body} some numerical results for different
Skyrme interactions up to N3LO level (as well as for others discussed later in the article). 
Along this article we have used the value $\hbar^2/2m = 20.735$~MeV fm$^2$. 
Notice that this value may not exactly be the one used by the authors of the various effective interactions 
mentioned here, but the possible differences are irrelevant for the present discussion.
For some forces we have checked that these differences are less than $5 \cdot 10^{-3}$\% and in all cases 
do not affect the numbers given in~Table~I.    

The selected interactions are the following:
\begin{itemize}
    \item At N1LO level, the two parametrisations SV~\cite{bei75} and SHZ2~\cite{sat12}, which do not contain any density dependent terms;
    \item Still at the N1LO level, a panel of various parametrisations  which differ mainly for their density-dependent term and for their 
				  values of the compression modulus. 
          Namely SIII~\cite{bei75}, SkM$^*$~\cite{bar82}, SLy5~\cite{cha97,cha98}, BSk1~\cite{sam02} and SLy5$^*$~\cite{ale13}.
    \item The N$\ell$LO parametrisation ($\ell=2,3$) obtained~\cite{dany14} using the Landau parameters derived from the finite-range 
				  interactions D1MT~\cite{gor09,ang12} and  M3Y-P2~\cite{nak03}, and hereafter called D1MT-N$\ell$LO and M3Y-P2-N$\ell$LO.
		\item The recent SN2LO1~\cite{pierre17} parametrisation built up
					via a complete minimisation of the penalty function based on both SNM properties and finite-nuclei observables.
\end{itemize}
										
Table~I 
gives the isoscalar bulk properties of each parametrisation
(first four columns).
The second part of this Table (next three columns) gives the two-body contribution to the isoscalar
effective mass as given by Eq.~(\ref{rel-N3LO}) --the $C_1^{(6)}$ term is denoted as $\Delta_{FR}$ for reasons 
which will be clarified below-- as well as the total isoscalar effective mass for the 
two body part only of the interactions. Keep in mind that the partial contributions concern 
$m/m^*$, {\it i.e.} the inverse of the effective mass.
 
The five first interactions listed on Table I are pure two-body
interactions. Therefore, the values for their effective masses given
by equation (24) match the full ones given by equation (41). These
values are slightly different of 0.4 because of the unusual saturation
densities and compression moduli predicted by some of these interactions.
For all other cases listed in Table I, the contribution to the effective
mass from the two-body part of the interaction as given by equation (24)
is close to 0.4 and therefore in agreement with Weisskopf's estimate.
In the case of N3LO interactions, one can see that the contribution of
the $C_1^{(6)}$ term is actually very small.
To get an estimate of the relative importance of such  terms entering Eq.~(\ref{rel-N3LO}), 
we take as an example the case of the M3Y-P2-N3LO interaction: 
we  get $1.375+1.0353+0.0126$, which leads to an the effective mass 0.4127. 
Dropping the last term, one gets 0.4149 instead. 
In conclusion, neglecting the $C_1^{(6)}$ term in Eq.~(\ref{rel-N3LO}), results in an overestimate 
of $m^*/m$ by less than $0.5\%$. 
Keeping this result in mind, we will now see how these results are modified with an explicit
finite-range interaction. 

%
%
\begin{table*}[ht!]
\begin{center}
\label{tab:2body}
\caption{Properties of the 2-body interactions used in this study at saturation density.
The various contributions to the isoscalar effective mass are given (see text for details).
All the finite range contributions given in the $\Delta_{FR}$ column are from finite
range interactions, $i.e.$ Gogny or Nakada interactions, except for the
zero range Skyrme-like DM1T and M3Y interactions where this column gives the N3LO contribution.
All the density dependencies are zero-range, $i.e.$ $t_3$-term, except for the D2 Gogny force
which uses a finite range density dependence with a Gaussian form factor. 
Missing entries are zero.} 
\smallskip 
\begin{tabular}{l cc cc cc cc cc cc cc cc cc cc cc cc}
\hline \hline \noalign{\smallskip}
 && $\rho_{sat}$ && $k_F$ && ${\cal E}_0$ && $K_\infty$ 
 && \vline && $\frac{5}{72} \left( K_\infty - 21 {\cal E}_0 \right)/\eps_F$ && $\Delta_{FR}$ && $m^*/m$  
 && \vline && $\alpha$ && $t_3$ && $m^*/m$ \\
 && (fm$^{-3}$) && (fm$^{-1}$) && (MeV) && (MeV) && \vline &&  && &&  Eq.~(\ref{rel-N3LO}) && \vline && &&  && Eq.~(\ref{rel-2B-DD}) \\
\noalign{\smallskip} \hline \noalign{\smallskip}
SV~\cite{bei75}   && 0.155 && 1.319 && -16.05 && 306 && \vline && 1.237 && && 0.383    
                                                     && \vline && && && 0.383  \\[0.3mm]
SHZ2~\cite{sat12} && 0.157 && 1.326 && -16.27 && 310 && \vline && 1.241 && && 0.382 
																									   && \vline && && && 0.382  \\[0.3mm]
                  &&       &&       &&        &&     && \vline && && &&         
                                                     && \vline && && &&        \\[0.3mm]
B1~\cite{bri67}   && 0.205 && 1.448 && -15.69 && 183 && \vline && 0.819 && 0.0225 && 0.451  
                                                     && \vline && && && 0.451  \\[0.3mm]
C1~\cite{bri67}   && 0.206 && 1.451 && -15.83 && 218 && \vline && 0.876 && 0.0133 && 0.442 
                                                     && \vline && && && 0.442  \\[0.3mm]
L3~\cite{bri67}   && 0.277 && 1.601 && -15.75 && 216 && \vline && 0.714 && 0.0090 && 0.477 
                                                     && \vline && && && 0.477  \\[0.3mm] 
                  &&       &&       &&        &&     && \vline && && &&       
                                                     && \vline && && &&        \\[0.3mm]
SIII~\cite{bei75}    && 0.145 && 1.291 && -15.85 && 355 && \vline && 1.383 &&        && 0.363 
                                                        && \vline && 1     && -1.448 && 0.763 \\[0.3mm]
SkM$^*$~\cite{bar82} && 0.160 && 1.334 && -15.77 && 217 && \vline && 1.031 && && 0.416 
                                                        && \vline && $\frac{1}{6}$ && -1.138 && 0.789 \\[0.3mm]
BSk1~\cite{sam02}    && 0.157 && 1.325 && -15.80 && 231 && \vline && 1.074 && && 0.408 
                                                        && \vline && $\frac{1}{3}$ && -1.496 && 1.050 \\[0.3mm]
SLy5~\cite{cha97,cha98} && 0.160 && 1.334 && -15.98 && 230 && \vline && 1.0644 && && 0.410 
                                                           && \vline && $\frac{1}{6}$ && -1.005 && 0.697 \\[0.3mm]
SLy5$^*$~\cite{ale13}   && 0.161 && 1.334 && -16.02 && 230 && \vline && 1.065 && && 0.410 
                                                           && \vline && $\frac{1}{6}$ && -1.013 && 0.701 \\[0.3mm]								
M3Y-P2-N1LO~\cite{dany14} && 0.162 && 1.338 && -12.35 && 217 && \vline && 0.890 && && 0.441      
                                                             && \vline && $\frac{1}{3}$ && -0.730 && 0.652 \\[0.3mm]
                          &&       &&       &&        &&     && \vline &&       && &&       
                                                             && \vline && && &&        \\[0.3mm]
SN2LO1~\cite{pierre17}     && 0.162 && 1.339 && -15.95 && 222 && \vline && 1.041 && && 0.414 
                                                              && \vline && $\frac{1}{6}$ && -1.005 && 0.709 \\[0.3mm]
DM1T-N2LO~\cite{dany14}    && 0.143 && 1.284 &&  -9.92 && 154 && \vline && 0.736 && && 0.474       
                                                              && \vline && $\frac{1}{3}$ && -0.773 && 0.748 \\[0.3mm]
M3Y-P2-N2LO~\cite{dany14}  && 0.158 && 1.327 && -15.25 && 206 && \vline && 1.001 && && 0.421       
                                                              && \vline && $\frac{1}{3}$ && -0.839 && 0.651 \\[0.3mm]
                           &&       &&       &&        &&     && \vline &&       && &&       
                                                              && \vline &&       &&        &&       \\[0.3mm]
DM1T-N3LO~\cite{dany14}    && 0.164 && 1.345 && -15.35 && 215 && \vline && 0.996 && 0.0028 && 0.422      
                                                              && \vline && $\frac{1}{3}$ && -1.044 && 0.776 \\[0.3mm]
M3Y-P2-N3LO~\cite{dany14}  && 0.161 && 1.337 && -15.96 && 217 && \vline && 1.035 && 0.0126 && 0.413      
                                                              && \vline && $\frac{1}{3}$ && -0.886 && 0.651 \\[0.3mm]
                           &&       &&       &&        &&     && \vline &&       && &&       
                                                              && \vline &&    &&        &&       \\[0.3mm]
D1~\cite{d1}   && 0.166 && 1.351  && -16.30 && 229 && \vline && 1.050 && 0.0047 && 0.412        
                                                   && \vline && $\frac{1}{3}$ && -0.936 && 0.670 \\[0.3mm]
D1S~\cite{d1s} && 0.163 && 1.342  && -16.01 && 203 && \vline && 1.002 && 0.0029 && 0.419       
                                                   && \vline && $\frac{1}{3}$ && -0.952 && 0.697 \\[0.3mm]
D1M~\cite{gor09} && 0.165 && 1.346  && -16.02 && 225 && \vline && 1.038 && 0.0030 && 0.414        
                                                   && \vline && $\frac{1}{3}$ && -1.076 && 0.746 \\[0.3mm]
					  	 &&        &&       &&        &&     && \vline &&       && &&      
                                                   && \vline &&       && &&                      \\[0.3mm]	
D2~\cite{d2}   && 0.161 && 1.337  && -15.82 && 207 && \vline && 1.011 && 0.0170 && 0.416       
                                                   && \vline && $\frac{1}{3}$ && -1.047 && 0.738 \\[0.3mm]
               &&        &&       &&        &&     && \vline &&       && &&       
																									 && \vline &&       && &&                      \\[0.3mm]										
P2~\cite{nak03} && 0.163 && 1.340 && -16.14 && 220 && \vline && 1.043 && 0.0167 && 0.411    
                                                   && \vline && $\frac{1}{3}$ && -0.901 && 0.652 \\[0.3mm]
P6~\cite{nak13} && 0.163 && 1.340 && -16.24 && 240 && \vline && 1.083 && 0.0161 && 0.404
                                                   && \vline && $\frac{1}{3}$ && -0.797 && 0.596 \\[0.3mm]
P7~\cite{nak13} && 0.163 && 1.340 && -16.23 && 255 && \vline && 1.111 && 0.0158 && 0.400
                                                   && \vline && $\frac{1}{3}$ && -0.803 && 0.589 \\[0.3mm]
\noalign{\smallskip} \hline \hline
\end{tabular}
\end{center}
\end{table*}
%
%
%

%
%
\subsection{Finite-range interaction}
\label{sec:finite2B}
%
%

To be as general as possible, we consider a finite-range two-body interaction written as

\be
V = \sum_n\left( W_n+B_n P_{\sigma} - H_n P_{\tau} - M_n P_{\sigma \tau} \right) V(r/\mu_n)\;.
\label{eq:bb}
\ee
The radial form factor is characterized by a set of ranges denoted $\mu_n$.
For the sake of simplicity, the index $n$ will be omitted in the following; 
a sum over it is to be understood in every term containing a range $\mu$.
From this interaction, we derive the energy per particle
\bea
\frac{E}{A} & = & \frac{3}{5} \, \frac{\hbar^2}{2 m} \, k_F^2 
           +  2  \pi  \rho \, C_D \, \int \, {\rm d}r \, r^2 \, V(r/\mu)  \nnn
            &   & - \frac{12}{\pi} C_E  \, k_F^3 \, \int \, {\rm d}r\, r^2  V(r/\mu)
                                           \, \left[ \frac{j_1(k_F r)}{k_F r}\right]^2 ,
\eea
where
\bea
C_D &=& W + \frac{1}{2} B - \frac{1}{2} H - \frac{1}{4} M \;, \\
C_E &=& \frac{1}{4} W + \frac{1}{2} B - \frac{1}{2} H -  M \;, 
\eea 
are respectively the combinations related to the direct and exchange contributions. 
As already mentioned, we see from the above equation that the direct contribution is linear in $\rho_0$.

As for the zero-range case, we determine the quantities of interest. 
The starting expressions are
\bea
{\cal E}_0 &=& \frac{1}{5} \varepsilon_F  \nnn
&-& \frac{12}{\pi} C_E k_F^3 \int {\rm d}r \, r^2 V(r/\mu)  {\cal F}^{\cal E}(k_F r)\,, \\
K_{\infty}  &=& -  \frac{6}{5} \varepsilon_F   \nnn
&-& \frac{12}{\pi} C_E k_F^3 \int {\rm d}r \,  r^2 V(r/\mu)   {\cal F}^{K}(k_F r)\,,   \\
\frac{\hbar^2}{2 m^*} k_F^2 &=& \varepsilon_F \nnn
&+&  \frac{12}{\pi} C_E k_F^3  \int {\rm d}r \, r^2 \,  V(r/\mu)   {\cal F}^{m}(k_F r)\,, 
\eea
where we have defined the following functions
\bea
 {\cal F}^{\cal E}(x) &=& \frac{2}{x^2} j_1^2(x) - \frac{2}{3 x} j_0(x)  j_1(x)\,, \\
 {\cal F}^{K}(x) &=& 2 j_0^2(x) - \frac{12}{x} j_0(x) j_1(x) \nnn
 && +  \left(\frac{18}{x^2}- 2 \right) j_1^2(x)\,, \\
 {\cal F}^{m}(x) &=& \frac{1}{3}  j_1^2(x)\,.  
\eea
Taking the linear combination given by Eq.~(\ref{magic}), we deduce the following relation (with $x=k_Fr$)
\bea
\frac{m}{m^*} & = & \frac{11}{8} \, + \, \frac{5}{72} \,  \frac{ K_{\infty} - 21 {\cal E}_0}{\varepsilon_F} \nnn
&&+ 
  \frac{12}{\pi} \, \frac{C_E}{\varepsilon_F} \, \int {\rm d}x\, x^2 \, V\left(\frac{x}{k_F \mu}\right) \nnn
&&  \bigg\{  {\cal F}^{m}(x) + \frac{5}{72}  {\cal F}^{K}(x)
    - \frac{105}{72} {\cal F}^{\cal E}(x) \bigg\}\,. 
\label{rel-finite}
\eea
As expected, there is no contribution from the direct term $C_D$. 
On the contrary, the term containing the $C_E$ coefficient contains all finite-range effects, which go beyond 
the $C_1^{(6)}$ term of the N3LO equivalent relation given in Eq.~(\ref{rel-N3LO}). 
As long as the form factor $V(x)$ is short-ranged, it seems reasonable to take the following power expansion:
\bea
&& {\cal F}^{m}(x) + \frac{5}{72}  {\cal F}^{\cal K}(x) -\frac{105}{72}  {\cal F}^{\cal E}(x)  \nnn
 && \simeq  \frac{1}{127575} x^6 - \frac{8}{9823275} x^8 + 
\frac{1}{25540515} x^{10} +  \dots
\nn
\eea
We see that $x^2$ and $x^4$ contributions are exactly cancelled out, in agreement with our previous discussion. 
The first $x^6$ contribution is thus related to N3LO and the next contributions $x^n\;\; (n = 8, ...)$ to higher orders.

The contribution of the exchange term $C_E$ entering Eq.~(\ref{rel-finite}) will be denoted as $\Delta_{FR}$. 
It contains the explicit finite-range contribution to $m/m^*$. 

In Table~I 
are displayed results for several finite-range interactions. 
We have selected the following

\begin{itemize}
               \item The Brink-Boeker~\cite{bri67} interactions B1, C1 and L3 which use Gaussian form factors
                     and do not contain any density dependencies.
               \item The standard D1~\cite{d1} and D1S~\cite{d1s} Gogny interactions which use two
                     Gaussian form factors plus a zero range density dependence.
                     The D1M parametrisation~\cite{gor09} used in a first attempt to reproduce more 
										 than 2100 measured masses through a Gogny-Hartree-Fock-Bogoliubov Nuclear Mass Model.                   
                     The recent D2~\cite{d2} parametrisation which uses a finite range density dependence 
                     with a Gaussian form factor.
               \item Three parametrisations based on three Yukawa form factors~\cite{nak03,nak13} which also 
                     include explicit density dependent terms.
\end{itemize}               

One can see that $\Delta_{FR}$ is very small. 
To get an estimate of the relative importance of the terms entering Eq.~(\ref{rel-finite}), we give 
precisely the numerical values obtained for the D1 interaction: $1.375+1.0495+0.00469$, which result in the 
effective mass 0.4116. 
Dropping the last term, one gets 0.4125 instead which gives an overestimate less than $\approx 0.5\%$. 
For the Nakada's series the overestimate is larger, $i.e.$ around $1.6\%$ and this is due to the longer range
of the form factors used.

%
%
\subsection{Summary concerning the two-body contributions}
%
%

Either for zero- or finite-range effective interactions, one can write

\be
\frac{m}{m^*}\bigg|_{2B} \, = \, \frac{11}{8} 
                         \, + \, \frac{5}{72} \, \frac{1}{\varepsilon_F} \, \left( K_{\infty} - 21 {\cal E}_0 \right)     + \Delta_{\rm FR} \;,
\label{rel-2B}
\ee
where $\Delta_{FR}$ is a short-notation for the last terms entering Eqs.~(\ref{rel-N3LO}) and (\ref{rel-finite}) and it takes into account finite-range effects beyond N2LO.
In all cases examined here, it turns out that $\Delta_{\rm FR}$ is very small. 
The value $m^*/m \simeq 0.4$ is obtained for {\it any} two-body interaction giving reasonable 
${\cal E}_0$, $K_{\infty}$ and $k_F$ values. The same value is obtained when using only the pure two-body part of any of the effective interactions considered. This generalizes the older result obtained by Weisskopf.

Since the empirical effective mass in the bulk is expected to be around 0.7~\cite{zuo99,ma04}, other contributions
beyond the two-body terms are needed and mainly justify the use of a density dependence to
simulate such effects. The next part is devoted to that topic.

%
%
\section{Effective two-body density dependent interaction}
\label{sec:DD}
%
%

Since an effective interaction limited to a purely two-body term can not lead to a reasonable value 
of the effective mass, the inclusion of three- or even four-body terms seems to be unavoidable. 
However, to the best of  our knowledge, even if some pioneering work has been done in a recent past~\cite{sad13,oni78,sad13b,ben17}, 
decisive improvements remain necessary to use such interactions for practical applications.
For these reasons, and because we are mainly interested in the properties of infinite nuclear matter at the mean-field level 
in this article, in this section we limit ourselves to the usual effective two-body 
density-dependent term which is currently used either in zero- or finite-range interactions. 
Explicitly, this term reads
\be
V_{DD} = \frac{1}{6} t_3 (1 + x_3 P_{\sigma}) \rho^{\alpha}\,.
\label{t3}
\ee
Let us remark that the factor 1/6 reflects the historical three-body origin of this term~\cite{vau72}. 
This coefficient is usually used for zero range Skyrme interaction, but disregarded for 
the finite range one.

For zero-range interactions, this term does not directly contribute to the effective mass, 
while its contribution to the other quantities entering  Eq.~(\ref{magic}) is
\bea
\label{e0-dd}
\left. {\cal E}_0  \right|_{DD}
          &=& - \frac{3}{8} \frac{1}{6} t_3 \rho_0^{\alpha+1} \alpha \\
\label{kinf-dd}
\left. K_{\infty}  \right|_{DD}
         &=&  \frac{27}{8} \frac{1}{6} t_3 \rho_0^{\alpha+1} \alpha (\alpha+1)\,.
\eea
It is worth mentioning at this stage that, in the N1LO case, the parameter
$t_3$ is sometimes eliminated with an appropriate combination of ${\cal E}_0$
and $K_{\infty}$, namely
\begin{align}
\label{eq:Kmstar}
9(\alpha+1) {\cal E}_0 + K_{\infty} =& 
\frac{3}{5} (3 \alpha+1) \varepsilon_F \nonumber \\
& - \frac{3}{40} (3 \alpha-2) C_1^{(2)} \rho_0 k_F^2\,.
\end{align}
The value $\alpha=2/3$ leads to a singular situation specific to the standard Skyrme interaction 
for which $K_\infty$ is determined by ${\cal E}_0$ and $k_F$ and is not related to the effective mass. 
However, since we are interested in getting general relations valid for any effective interaction, 
either zero-or finite-range, we will consider in the following the complete equations, that is beyond N1LO. 
Therefore, plugging the density-dependent contributions~(\ref{e0-dd}) and~(\ref{kinf-dd}) into Eq.~(\ref{magic}) 
we get
\bea
\frac{m}{m^*} &=&  \frac{11}{8} + \frac{5}{72} \frac{K_{\infty}
                   - 21 {\cal E}_0}{\varepsilon_F} 
                   + \Delta_{\rm FR} \nnn
&& -  \frac{5}{384} \alpha ( 10 + 3 \alpha)
               \frac{t_3 \rho_0^{\alpha+1}}{\varepsilon_F}\,.
\label{rel-2B-DD}
\eea
In Table~I 
are collected the contributions of all the
terms entering Eq.~(\ref{rel-2B-DD}), but the constant 11/8. 

The conclusions of this analysis are clear:

\begin{itemize}
               \item [$\bullet$] the pure two-body part of any interaction leads to an effective mass value is solely determined 
                                 by the values of ${\cal E}_0, K_{\infty}, k_F$. When the accepted values of these parameters are used, one gets $m^*/m \simeq 0.4$, within less that 1\% accuracy.
               \item [$\bullet$] the density-dependent term substantially modifies that value, increasing it to typically $m^*/m \simeq 0.7$.
\end{itemize}

%
%
%
%

\begin{figure}[h!]
\begin{center}
\includegraphics[width=0.9\linewidth,clip=true]{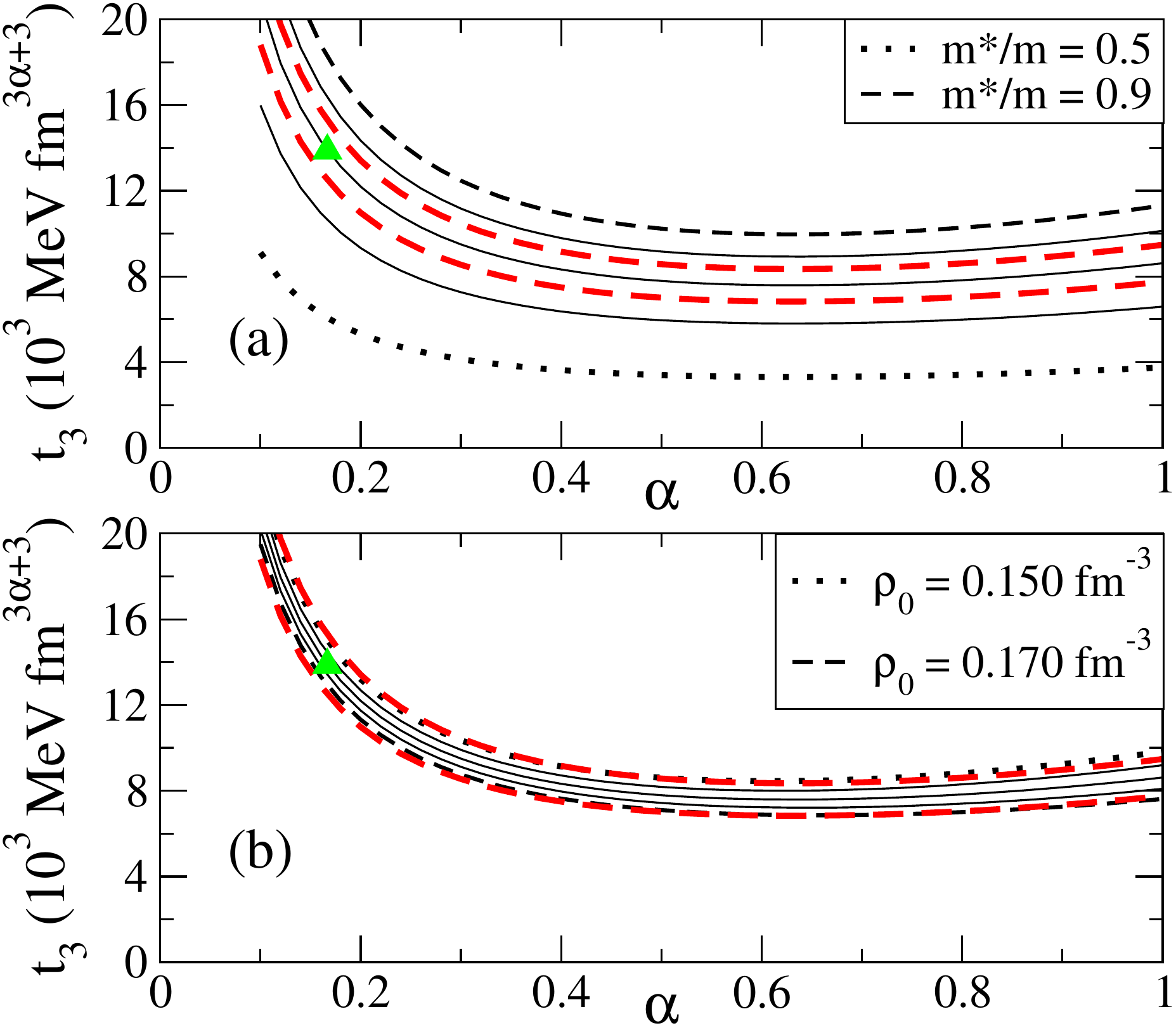}
\includegraphics[width=0.9\linewidth,clip=true]{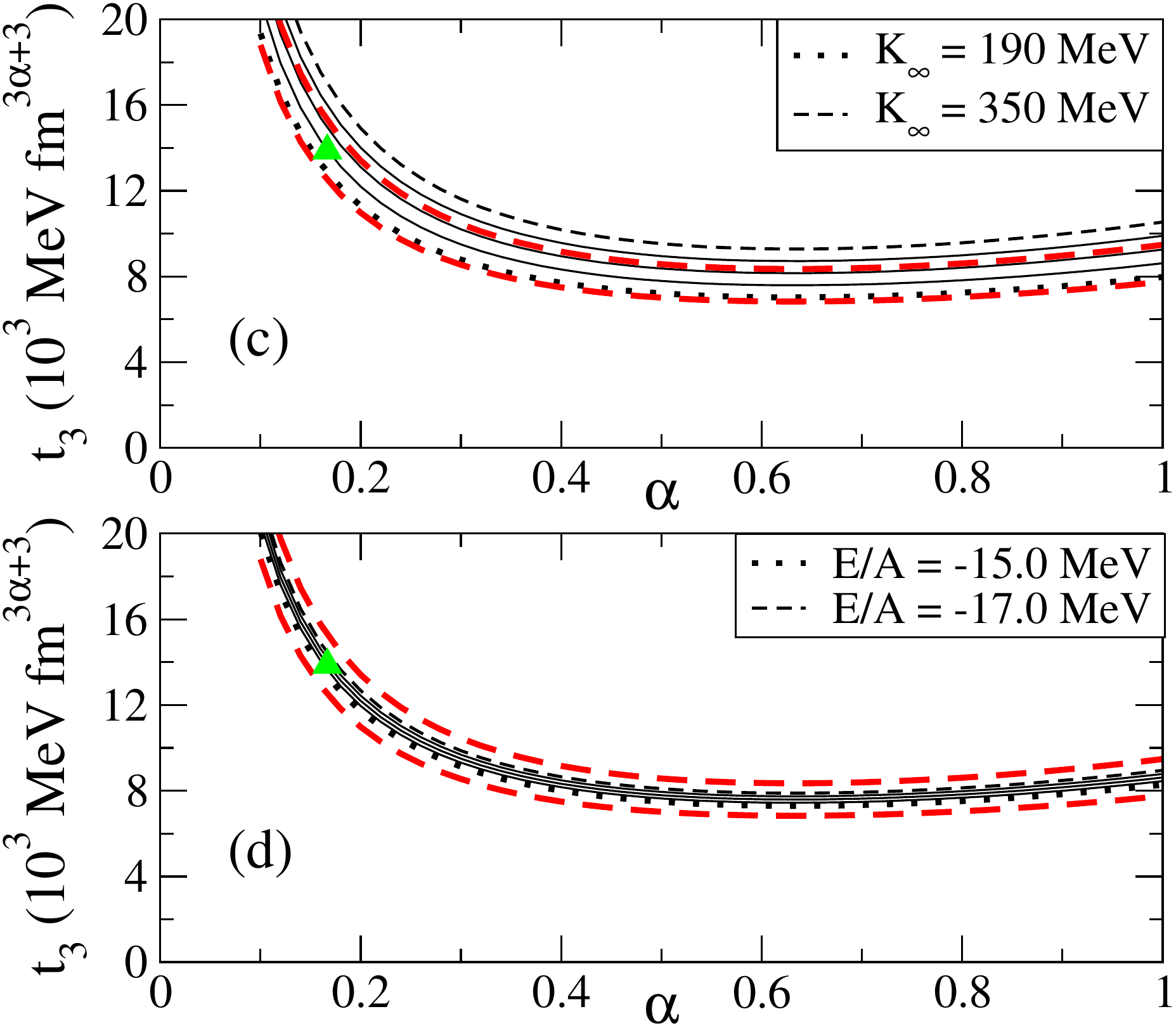}
\end{center}
\caption{(Color online) The parameter $t_3$ plotted as a function of $\alpha$ for various sets of isoscalar bulk properties.
In each case, one of the four bulk properties $K_\infty, m^*/m, E/A, \rhosat$ is varied
keeping constant the three others (see text). 
The green triangle marks the SLy5$^*$ parametrisation~\cite{ale13}.}
\label{fig:t3_alpha}
\end{figure}

An important remark concerning Eq.~(\ref{rel-2B-DD}) is that it actually provides an unexpected relation between the parameters $t_3$ and $\alpha$. For (reasonable) fixed values of inputs ${\cal E}_0$, $K_{\infty}$, $k_F$, these parameters are not independent. 
This is reflected in Fig.~\ref{fig:t3_alpha} for several sets of these quantities. Actually, we have dropped the $\Delta_{FR}$ term in these figures. 
We start from the set $K_\infty=230 \pm 20$~MeV, $m^*/m=0.70 \pm 0.02$, ${\cal E}_0=-16.0 \pm 0.5$~MeV and $\rho_{0}=0.160 \pm 0.005$~fm$^{-3}$. The two thick dashed lines in all panels represent the limits of $t_3$ as a function of $\alpha$ taking into account, in a schematic way, the statistical errors related to the uncertainties of the inputs as well as the neglected small contribution from $\Delta_{FR}$. 
The remaining curves correspond to the central values of three of these inputs, for varying values of the fourth one.   
In panel (a) the effective mass is varied in steps of 0.1; in panel (b) the density is varied in steps of 0.05~fm$^{-3}$; in panel (c) the compression modulus is varied in steps of 40~MeV; in panel (d) the energy per particle is varied in steps of 0.5~MeV. One can see that the relation between $t_3$ and $\alpha$ is not very sensitive to the inputs for $\rho_0$ and ${\cal E}_0$, but to the (extreme) values for the effective mass.

%
%
%

For values of $\alpha$ in the interval between 0.4 and 0.9, the coefficient $t_3$ has an almost constant value (around 8000, in the right units), with a minimum for $\alpha \simeq 0.64$. All the curves show a divergence of the $t_3$ parameter when $\alpha$ decreases showing
some dangerous range of values smaller than 1/6. Here, we see the main reason why it is dangerous to include the $\alpha$ parameter 
in the fitting procedure.

\section{Conclusions}
\label{sec:conclusion}
%
%

In this article, we have performed a systematic study on the properties of the effective mass for general two-body interactions, 
both zero- and finite-range.
In particular, starting from the previous work of Weisskopf~\cite{wei57}, we have shown that the two-body part of $any$ effective interaction induces an effective mass at most of  0.4, irrespectively of the range of the interaction, as far as other infinite matter properties are kept to reasonable values.
The result is exact for N1LO (Skyrme) and N2LO interactions, while there is a minor correction of $\approx 1-2$\% for the higher order N3LO pseudo-potential and/or finite range interactions.

To increase the value of the effective mass to higher values without spoiling other infinite matter properties as saturation density and incompressibility, it is thus necessary to add either an explicit three-body term or equivalently, but more phenomenological, a two-body 
density dependent term.
The latter is the common strategy used by the vast majority of effective interaction available nowadays.

An interesting result of our analysis is the strong built-in correlation found between the intensity of the density dependent interaction 
$t_3$ and the exponent of the density $\alpha$. 
This strong correlation is only marginally affected by the explicit presence of a finite range or equivalently higher order gradient terms. This correlation reflects a lack of flexibility in our models and in particular in the way three-body terms are treated.

One should keep in mind that the standard density-dependent term was originally generated~\cite{vau72} from a simple zero-range three-body
interaction. Including a more general three-body interaction seems to be the proper way to go beyond a pure two-body interaction. Among the several attempts in this direction, we just mention two of them: 
a zero-range three-body interactions including gradient terms ~\cite{sad13b,sad13} and a semicontact three-body interaction~\cite{lac15}. In both cases we have verified that one gets the expected result, namely the pure two-body part gives a contribution to the effective mass of about 0.4, while the three-body part  increases this value to the accepted one. It is possible to get for these interactions a relation  similar to Eq.~(\ref{rel-2B-DD}), in which the density-dependent contribution is replaced with a three-body contribution. Interestingly, this equation provides a  correlation between three-body parameters, analogous to that between $t_3$ and $\alpha$, which could be usefully utilized in the process of determining the parameters.

%
%
\section*{Acknowlegments}
We thank G. Bertsch for drawing our attention to ref.~\cite{wei57} and M. Bender for very fruitful discussions.
JN is supported by grant FIS2017-84038-C2-1-P, Mineco (Spain).
  The work of AP is supported  by the UK Science and Technology Facilities Council under Grants No. ST/L005727 and ST/M006433.   
 
%
%

%
%
%
%
%
%


%
%

%
%

%
%


%
\bibliography{effmass}

\begin{thebibliography}{30}
\expandafter\ifx\csname natexlab\endcsname\relax\def\natexlab#1{#1}\fi
\expandafter\ifx\csname bibnamefont\endcsname\relax
  \def\bibnamefont#1{#1}\fi
\expandafter\ifx\csname bibfnamefont\endcsname\relax
  \def\bibfnamefont#1{#1}\fi
\expandafter\ifx\csname citenamefont\endcsname\relax
  \def\citenamefont#1{#1}\fi
\expandafter\ifx\csname url\endcsname\relax
  \def\url#1{\texttt{#1}}\fi
\expandafter\ifx\csname urlprefix\endcsname\relax\def\urlprefix{URL }\fi
\providecommand{\bibinfo}[2]{#2}
\providecommand{\eprint}[2][]{\url{#2}}

\bibitem[{\citenamefont{Chabanat et~al.}(1997)\citenamefont{Chabanat, Bonche,
  Haensel, Meyer, and Schaeffer}}]{cha97}
\bibinfo{author}{\bibfnamefont{E.}~\bibnamefont{Chabanat}},
  \bibinfo{author}{\bibfnamefont{P.}~\bibnamefont{Bonche}},
  \bibinfo{author}{\bibfnamefont{P.}~\bibnamefont{Haensel}},
  \bibinfo{author}{\bibfnamefont{J.}~\bibnamefont{Meyer}}, \bibnamefont{and}
  \bibinfo{author}{\bibfnamefont{R.}~\bibnamefont{Schaeffer}},
  \bibinfo{journal}{Nucl. Phys. A} \textbf{\bibinfo{volume}{627}},
  \bibinfo{pages}{710 } (\bibinfo{year}{1997}).

\bibitem[{\citenamefont{Beiner et~al.}(1975)\citenamefont{Beiner, Flocard,
  Giai, and Quentin}}]{bei75}
\bibinfo{author}{\bibfnamefont{M.}~\bibnamefont{Beiner}},
  \bibinfo{author}{\bibfnamefont{H.}~\bibnamefont{Flocard}},
  \bibinfo{author}{\bibfnamefont{N.~V.} \bibnamefont{Giai}}, \bibnamefont{and}
  \bibinfo{author}{\bibfnamefont{P.}~\bibnamefont{Quentin}},
  \bibinfo{journal}{Nucl. Phys. A} \textbf{\bibinfo{volume}{238}},
  \bibinfo{pages}{29 } (\bibinfo{year}{1975}).

\bibitem[{\citenamefont{Satu\l{}a et~al.}(2012)\citenamefont{Satu\l{}a,
  Dobaczewski, Nazarewicz, and Werner}}]{sat12}
\bibinfo{author}{\bibfnamefont{W.}~\bibnamefont{Satu\l{}a}},
  \bibinfo{author}{\bibfnamefont{J.}~\bibnamefont{Dobaczewski}},
  \bibinfo{author}{\bibfnamefont{W.}~\bibnamefont{Nazarewicz}},
  \bibnamefont{and} \bibinfo{author}{\bibfnamefont{T.~R.}
  \bibnamefont{Werner}}, \bibinfo{journal}{Phys. Rev. C}
  \textbf{\bibinfo{volume}{86}}, \bibinfo{pages}{054316}
  (\bibinfo{year}{2012}).

\bibitem[{\citenamefont{Brink and Boeker}(1967)}]{bri67}
\bibinfo{author}{\bibfnamefont{D.}~\bibnamefont{Brink}} \bibnamefont{and}
  \bibinfo{author}{\bibfnamefont{E.}~\bibnamefont{Boeker}},
  \bibinfo{journal}{Nucl. Phys. A} \textbf{\bibinfo{volume}{91}},
  \bibinfo{pages}{1 } (\bibinfo{year}{1967}).

\bibitem[{\citenamefont{Bennaceur et~al.}(2017)\citenamefont{Bennaceur, Idini,
  Dobaczewski, Dobaczewski, Kortelainen, and Raimondi}}]{ben17}
\bibinfo{author}{\bibfnamefont{K.}~\bibnamefont{Bennaceur}},
  \bibinfo{author}{\bibfnamefont{A.}~\bibnamefont{Idini}},
  \bibinfo{author}{\bibfnamefont{J.}~\bibnamefont{Dobaczewski}},
  \bibinfo{author}{\bibfnamefont{P.}~\bibnamefont{Dobaczewski}},
  \bibinfo{author}{\bibfnamefont{M.}~\bibnamefont{Kortelainen}},
  \bibnamefont{and} \bibinfo{author}{\bibfnamefont{F.}~\bibnamefont{Raimondi}},
  \bibinfo{journal}{J. Phys. G: Nucl. Part. Phys.}
  \textbf{\bibinfo{volume}{44}}, \bibinfo{pages}{045106}
  (\bibinfo{year}{2017}).

\bibitem[{\citenamefont{Weisskopf}(1957)}]{wei57}
\bibinfo{author}{\bibfnamefont{V.~F.} \bibnamefont{Weisskopf}},
  \bibinfo{journal}{Nucl. Phys.} \textbf{\bibinfo{volume}{3}},
  \bibinfo{pages}{423 } (\bibinfo{year}{1957}).

\bibitem[{\citenamefont{Nakatsukasa et~al.}(2016)\citenamefont{Nakatsukasa,
  Matsuyanagi, Matsuo, and Yabana}}]{nak16}
\bibinfo{author}{\bibfnamefont{T.}~\bibnamefont{Nakatsukasa}},
  \bibinfo{author}{\bibfnamefont{K.}~\bibnamefont{Matsuyanagi}},
  \bibinfo{author}{\bibfnamefont{M.}~\bibnamefont{Matsuo}}, \bibnamefont{and}
  \bibinfo{author}{\bibfnamefont{K.}~\bibnamefont{Yabana}},
  \bibinfo{journal}{Rev. Mod. Phys.} \textbf{\bibinfo{volume}{88}},
  \bibinfo{pages}{045004} (\bibinfo{year}{2016}).

\bibitem[{\citenamefont{Hugenholtz and Van~Hove}(1958)}]{hvh}
\bibinfo{author}{\bibfnamefont{N.~M.} \bibnamefont{Hugenholtz}}
  \bibnamefont{and} \bibinfo{author}{\bibfnamefont{L.}~\bibnamefont{Van~Hove}},
  \bibinfo{journal}{Physica} \textbf{\bibinfo{volume}{24}},
  \bibinfo{pages}{363} (\bibinfo{year}{1958}).

\bibitem[{\citenamefont{Raimondi et~al.}(2011)\citenamefont{Raimondi, Carlsson,
  and Dobaczewski}}]{rai11}
\bibinfo{author}{\bibfnamefont{F.}~\bibnamefont{Raimondi}},
  \bibinfo{author}{\bibfnamefont{B.~G.} \bibnamefont{Carlsson}},
  \bibnamefont{and}
  \bibinfo{author}{\bibfnamefont{J.}~\bibnamefont{Dobaczewski}},
  \bibinfo{journal}{Phys. Rev. C} \textbf{\bibinfo{volume}{83}},
  \bibinfo{pages}{054311} (\bibinfo{year}{2011}).

\bibitem[{\citenamefont{Davesne et~al.}(2014)\citenamefont{Davesne, Pastore,
  and Navarro}}]{dany14}
\bibinfo{author}{\bibfnamefont{D.}~\bibnamefont{Davesne}},
  \bibinfo{author}{\bibfnamefont{A.}~\bibnamefont{Pastore}}, \bibnamefont{and}
  \bibinfo{author}{\bibfnamefont{J.}~\bibnamefont{Navarro}},
  \bibinfo{journal}{J. Phys. G: Nucl. Part. Phys.}
  \textbf{\bibinfo{volume}{41}}, \bibinfo{pages}{065104}
  (\bibinfo{year}{2014}).

\bibitem[{\citenamefont{Davesne et~al.}(2016)\citenamefont{Davesne, Becker,
  Pastore, and Navarro}}]{dav16}
\bibinfo{author}{\bibfnamefont{D.}~\bibnamefont{Davesne}},
  \bibinfo{author}{\bibfnamefont{P.}~\bibnamefont{Becker}},
  \bibinfo{author}{\bibfnamefont{A.}~\bibnamefont{Pastore}}, \bibnamefont{and}
  \bibinfo{author}{\bibfnamefont{J.}~\bibnamefont{Navarro}},
  \bibinfo{journal}{Ann. Phys. (NY)} \textbf{\bibinfo{volume}{375}},
  \bibinfo{pages}{288} (\bibinfo{year}{2016}).

\bibitem[{\citenamefont{Bartel et~al.}(1982)\citenamefont{Bartel, Quentin,
  Brack, Guet, and H{\aa}kansson}}]{bar82}
\bibinfo{author}{\bibfnamefont{J.}~\bibnamefont{Bartel}},
  \bibinfo{author}{\bibfnamefont{P.}~\bibnamefont{Quentin}},
  \bibinfo{author}{\bibfnamefont{M.}~\bibnamefont{Brack}},
  \bibinfo{author}{\bibfnamefont{C.}~\bibnamefont{Guet}}, \bibnamefont{and}
  \bibinfo{author}{\bibfnamefont{H.-B.} \bibnamefont{H{\aa}kansson}},
  \bibinfo{journal}{Nucl. Phys. A} \textbf{\bibinfo{volume}{386}},
  \bibinfo{pages}{79 } (\bibinfo{year}{1982}).

\bibitem[{\citenamefont{Chabanat et~al.}(1998)\citenamefont{Chabanat, Bonche,
  Haensel, Meyer, and Schaeffer}}]{cha98}
\bibinfo{author}{\bibfnamefont{E.}~\bibnamefont{Chabanat}},
  \bibinfo{author}{\bibfnamefont{P.}~\bibnamefont{Bonche}},
  \bibinfo{author}{\bibfnamefont{P.}~\bibnamefont{Haensel}},
  \bibinfo{author}{\bibfnamefont{J.}~\bibnamefont{Meyer}}, \bibnamefont{and}
  \bibinfo{author}{\bibfnamefont{R.}~\bibnamefont{Schaeffer}},
  \bibinfo{journal}{Nucl. Phys. A} \textbf{\bibinfo{volume}{635}},
  \bibinfo{pages}{231 } (\bibinfo{year}{1998}).

\bibitem[{\citenamefont{Samyn et~al.}(2002)\citenamefont{Samyn, Goriely,
  Heenen, Pearson, and Tondeur}}]{sam02}
\bibinfo{author}{\bibfnamefont{M.}~\bibnamefont{Samyn}},
  \bibinfo{author}{\bibfnamefont{S.}~\bibnamefont{Goriely}},
  \bibinfo{author}{\bibfnamefont{P.-H.} \bibnamefont{Heenen}},
  \bibinfo{author}{\bibfnamefont{J.}~\bibnamefont{Pearson}}, \bibnamefont{and}
  \bibinfo{author}{\bibfnamefont{F.}~\bibnamefont{Tondeur}},
  \bibinfo{journal}{Nucl. Phys. A} \textbf{\bibinfo{volume}{700}},
  \bibinfo{pages}{142 } (\bibinfo{year}{2002}).

\bibitem[{\citenamefont{Pastore et~al.}(2013)\citenamefont{Pastore, Davesne,
  Bennaceur, Meyer, and Hellemans}}]{ale13}
\bibinfo{author}{\bibfnamefont{A.}~\bibnamefont{Pastore}},
  \bibinfo{author}{\bibfnamefont{D.}~\bibnamefont{Davesne}},
  \bibinfo{author}{\bibfnamefont{K.}~\bibnamefont{Bennaceur}},
  \bibinfo{author}{\bibfnamefont{J.}~\bibnamefont{Meyer}}, \bibnamefont{and}
  \bibinfo{author}{\bibfnamefont{V.}~\bibnamefont{Hellemans}},
  \bibinfo{journal}{Phys. Scr.} \textbf{\bibinfo{volume}{T154}},
  \bibinfo{pages}{014014} (\bibinfo{year}{2013}).

\bibitem[{\citenamefont{Goriely et~al.}(2009)\citenamefont{Goriely, Hilaire,
  Girod, and P\'eru}}]{gor09}
\bibinfo{author}{\bibfnamefont{S.}~\bibnamefont{Goriely}},
  \bibinfo{author}{\bibfnamefont{S.}~\bibnamefont{Hilaire}},
  \bibinfo{author}{\bibfnamefont{M.}~\bibnamefont{Girod}}, \bibnamefont{and}
  \bibinfo{author}{\bibfnamefont{S.}~\bibnamefont{P\'eru}},
  \bibinfo{journal}{Phys. Rev. Lett.} \textbf{\bibinfo{volume}{102}},
  \bibinfo{pages}{242501} (\bibinfo{year}{2009}).

\bibitem[{\citenamefont{Anguiano et~al.}(2012)\citenamefont{Anguiano, C\'o,
  Donno, and Lallena}}]{ang12}
\bibinfo{author}{\bibfnamefont{M.}~\bibnamefont{Anguiano}},
  \bibinfo{author}{\bibfnamefont{G.}~\bibnamefont{C\'o}},
  \bibinfo{author}{\bibfnamefont{V.~D.} \bibnamefont{Donno}}, \bibnamefont{and}
  \bibinfo{author}{\bibfnamefont{A.}~\bibnamefont{Lallena}},
  \bibinfo{journal}{Phys. Rev. C} \textbf{\bibinfo{volume}{86}},
  \bibinfo{pages}{054302} (\bibinfo{year}{2012}).

\bibitem[{\citenamefont{Nakada}(2003)}]{nak03}
\bibinfo{author}{\bibfnamefont{H.}~\bibnamefont{Nakada}},
  \bibinfo{journal}{Phys. Rev. C} \textbf{\bibinfo{volume}{68}},
  \bibinfo{pages}{014316} (\bibinfo{year}{2003}).

\bibitem[{\citenamefont{Becker et~al.}(2017)\citenamefont{Becker, Davesne,
  Meyer, Navarro, and Pastore}}]{pierre17}
\bibinfo{author}{\bibfnamefont{P.}~\bibnamefont{Becker}},
  \bibinfo{author}{\bibfnamefont{D.}~\bibnamefont{Davesne}},
  \bibinfo{author}{\bibfnamefont{J.}~\bibnamefont{Meyer}},
  \bibinfo{author}{\bibfnamefont{J.}~\bibnamefont{Navarro}}, \bibnamefont{and}
  \bibinfo{author}{\bibfnamefont{A.}~\bibnamefont{Pastore}},
  \bibinfo{journal}{Phys. Rev. C} \textbf{\bibinfo{volume}{96}},
  \bibinfo{pages}{044330} (\bibinfo{year}{2017}).

\bibitem[{\citenamefont{Decharg\'e and Gogny}(1980)}]{d1}
\bibinfo{author}{\bibfnamefont{J.}~\bibnamefont{Decharg\'e}} \bibnamefont{and}
  \bibinfo{author}{\bibfnamefont{D.}~\bibnamefont{Gogny}},
  \bibinfo{journal}{Phys. Rev. C} \textbf{\bibinfo{volume}{21}},
  \bibinfo{pages}{1568} (\bibinfo{year}{1980}).

\bibitem[{\citenamefont{Berger et~al.}(1991)\citenamefont{Berger, Girod, and
  Gogny}}]{d1s}
\bibinfo{author}{\bibfnamefont{J.}~\bibnamefont{Berger}},
  \bibinfo{author}{\bibfnamefont{M.}~\bibnamefont{Girod}}, \bibnamefont{and}
  \bibinfo{author}{\bibfnamefont{D.}~\bibnamefont{Gogny}},
  \bibinfo{journal}{Comp. Phys. Comm.} \textbf{\bibinfo{volume}{63}},
  \bibinfo{pages}{365 } (\bibinfo{year}{1991}).

\bibitem[{\citenamefont{Chappert et~al.}(2015)\citenamefont{Chappert, Pillet,
  Girod, and Berger}}]{d2}
\bibinfo{author}{\bibfnamefont{F.}~\bibnamefont{Chappert}},
  \bibinfo{author}{\bibfnamefont{N.}~\bibnamefont{Pillet}},
  \bibinfo{author}{\bibfnamefont{M.}~\bibnamefont{Girod}}, \bibnamefont{and}
  \bibinfo{author}{\bibfnamefont{J.-F.} \bibnamefont{Berger}},
  \bibinfo{journal}{Phys. Rev. C} \textbf{\bibinfo{volume}{91}},
  \bibinfo{pages}{034312} (\bibinfo{year}{2015}).

\bibitem[{\citenamefont{Nakada}(2013)}]{nak13}
\bibinfo{author}{\bibfnamefont{H.}~\bibnamefont{Nakada}},
  \bibinfo{journal}{Phys. Rev. C} \textbf{\bibinfo{volume}{87}},
  \bibinfo{pages}{014336} (\bibinfo{year}{2013}).

\bibitem[{\citenamefont{Zuo et~al.}(1999)\citenamefont{Zuo, Bombaci, and
  Lombardo}}]{zuo99}
\bibinfo{author}{\bibfnamefont{W.}~\bibnamefont{Zuo}},
  \bibinfo{author}{\bibfnamefont{I.}~\bibnamefont{Bombaci}}, \bibnamefont{and}
  \bibinfo{author}{\bibfnamefont{U.}~\bibnamefont{Lombardo}},
  \bibinfo{journal}{Physical Review C} \textbf{\bibinfo{volume}{60}},
  \bibinfo{pages}{024605} (\bibinfo{year}{1999}).

\bibitem[{\citenamefont{Ma et~al.}(2004)\citenamefont{Ma, Rong, Chen, Zhu, and
  Song}}]{ma04}
\bibinfo{author}{\bibfnamefont{Z.-Y.} \bibnamefont{Ma}},
  \bibinfo{author}{\bibfnamefont{J.}~\bibnamefont{Rong}},
  \bibinfo{author}{\bibfnamefont{B.-Q.} \bibnamefont{Chen}},
  \bibinfo{author}{\bibfnamefont{Z.-Y.} \bibnamefont{Zhu}}, \bibnamefont{and}
  \bibinfo{author}{\bibfnamefont{H.-Q.} \bibnamefont{Song}},
  \bibinfo{journal}{Physics Letters B} \textbf{\bibinfo{volume}{604}},
  \bibinfo{pages}{170} (\bibinfo{year}{2004}).

\bibitem[{\citenamefont{Sadoudi
  et~al.}(2013{\natexlab{a}})\citenamefont{Sadoudi, Bender, Bennaceur, Davesne,
  Jodon, and Duguet}}]{sad13}
\bibinfo{author}{\bibfnamefont{J.}~\bibnamefont{Sadoudi}},
  \bibinfo{author}{\bibfnamefont{M.}~\bibnamefont{Bender}},
  \bibinfo{author}{\bibfnamefont{K.}~\bibnamefont{Bennaceur}},
  \bibinfo{author}{\bibfnamefont{D.}~\bibnamefont{Davesne}},
  \bibinfo{author}{\bibfnamefont{R.}~\bibnamefont{Jodon}}, \bibnamefont{and}
  \bibinfo{author}{\bibfnamefont{T.}~\bibnamefont{Duguet}},
  \bibinfo{journal}{Phys. Scr.} \textbf{\bibinfo{volume}{T154}},
  \bibinfo{pages}{014013} (\bibinfo{year}{2013}{\natexlab{a}}).

\bibitem[{\citenamefont{Onishi and Negele}(1978)}]{oni78}
\bibinfo{author}{\bibfnamefont{N.}~\bibnamefont{Onishi}} \bibnamefont{and}
  \bibinfo{author}{\bibfnamefont{J.~W.} \bibnamefont{Negele}},
  \bibinfo{journal}{Nuclear Physics A} \textbf{\bibinfo{volume}{301}},
  \bibinfo{pages}{336} (\bibinfo{year}{1978}).

\bibitem[{\citenamefont{Sadoudi
  et~al.}(2013{\natexlab{b}})\citenamefont{Sadoudi, Duguet, Meyer, and
  Bender}}]{sad13b}
\bibinfo{author}{\bibfnamefont{J.}~\bibnamefont{Sadoudi}},
  \bibinfo{author}{\bibfnamefont{T.}~\bibnamefont{Duguet}},
  \bibinfo{author}{\bibfnamefont{J.}~\bibnamefont{Meyer}}, \bibnamefont{and}
  \bibinfo{author}{\bibfnamefont{M.}~\bibnamefont{Bender}},
  \bibinfo{journal}{Phys. Rev. C} \textbf{\bibinfo{volume}{88}},
  \bibinfo{pages}{064326} (\bibinfo{year}{2013}{\natexlab{b}}).

\bibitem[{\citenamefont{Vautherin and Brink}(1972)}]{vau72}
\bibinfo{author}{\bibfnamefont{D.}~\bibnamefont{Vautherin}} \bibnamefont{and}
  \bibinfo{author}{\bibfnamefont{D.}~\bibnamefont{Brink}},
  \bibinfo{journal}{Phys. Rev. C} \textbf{\bibinfo{volume}{5}},
  \bibinfo{pages}{626} (\bibinfo{year}{1972}).

\bibitem[{\citenamefont{Lacroix and Bennaceur}(2015)}]{lac15}
\bibinfo{author}{\bibfnamefont{D.}~\bibnamefont{Lacroix}} \bibnamefont{and}
  \bibinfo{author}{\bibfnamefont{K.}~\bibnamefont{Bennaceur}},
  \bibinfo{journal}{Physical Review C} \textbf{\bibinfo{volume}{91}},
  \bibinfo{pages}{011302} (\bibinfo{year}{2015}).

\end{thebibliography}
%
%
%
\end{document}